\documentclass[authorversion,pdftex, moreauthors, entropy, LaTeX, 10pt, a4paper]{mdpi} 

\firstpage{1} 
\copyrightyear{2023}
\pdfoutput=1 

\newcommand{\remove}[1]{}
\newcommand{\p}{\mathbf{p}}
\newcommand{\q}{\mathbf{q}}

\allowdisplaybreaks

\Title{Entropic Bounds on the Average Length  of Codes with a Space}

\Author{Roberto Bruno$^{\dagger}$\orcidA{} and Ugo Vaccaro*$^{,\dagger}$\orcidB{}} 

\AuthorNames{Roberto Bruno and Ugo Vaccaro}

\address[1]{%
Department of Computer Science, University of Salerno, 84084 Fisciano 
, Italy; rbruno@unisa.it}

\corres{Correspondence: uvaccaro@unisa.it
}

\secondnote{These authors contributed equally to this work.}

\abstract{{We consider the problem of constructing prefix-free codes 
in which a designated symbol, a \textit{space}, 
can only appear at the end of codewords.
We provide a linear-time algorithm to construct 
 \textit{almost}-optimal codes with this property, meaning that  
their average length differs from the \textit{minimum possible} by at most one.
We obtain our results by uncovering a relation between 
our class of codes and the class of one-to-one codes.
Additionally, we
derive upper and lower bounds to the average length 
of optimal prefix-free codes with a space in terms
of the source entropy.}}


\begin{document}

\section{Introduction}
Modern natural languages achieve the unique parsability of written
texts by inserting a special character
(i.e., a \textit{{space} 
}) between words
 \cite{DW} (See \cite{wiki} for a few exceptions to this rule).  
Classical Information Theory, instead, studies codes that achieve the
unique parsability of texts by imposing diverse combinatorial
properties on the codeword set: e.g., the prefix property,  unique decipherability, etc. \cite{CT}. With respect to
the \textit{efficiency} of such codes (usually measured via
the average number of code symbols {per}
source symbol), it is
well known that the Shannon entropy of the information source
constitutes a fundamental lower bound for it. 
On the other hand, if one drops the property of the
unique parsability of code messages into individual codewords,
and simply requires that different source symbols be encoded 
with different codewords, one can obtain codes (known as
\textit{one-to-one codes}) with efficiency
below the source Shannon entropy
(although not too much below; see, e.g., \cite{Alon,Blundo}).

Jaynes \cite{Jaynes} took the approach
of directly studying source codes in which
a designated character of
the code alphabet is \textit{exclusively} used as a word delimiter.
More precisely, Jaynes studied the possible decrease of the
noiseless channel capacity (see \cite{shannon}, p. 8)  
associated with any code that uses a 
designated symbol as an end-of-codeword mark, as compared with
the noiseless channel capacity of an unconstrained code.
Quite interestingly,
Jaynes proved that the 
decrease of the
noiseless channel capacity of codes with 
an end-of-codeword mark becomes negligible, as
the maximum codeword length increases. 

{In this paper, 
we study the problem of constructing prefix-free codes where a specific symbol
(referred to as a ‘space’) can only be positioned at the end of codewords. 
We refer to this kind of prefix code as \textit{prefix codes ending with a space}.
We 
develop a linear-time algorithm that  constructs
‘almost’-optimal codes with this characteristic, 
in the sense that the  average length 
of the constructed codes is 
at most one unit longer than the \textit{shortest possible} 
average length of any prefix-free code in which the space can appear 
only at the end of codewords. We prove this result by highlighting
a connection between our type of codes and the 
well-known class of one-to-one codes. We also provide upper and lower limits of the average length of optimal prefix codes ending with a space, expressed in 
terms of the source entropy and
the cardinality of the code alphabet.

The paper is structured as follows. In Section \ref{rel}, 
we illustrate the relationships between 
prefix codes ending with a space and one-to-one codes.
Specifically, we prove that, from one-to-one codes, one can easily
construct prefix codes ending with a space, and we give an upper bound on the 
average length of the constructed codes.
Successively, we show that, if we remove
 all the spaces from 
 the codewords of prefix  codes ending with a space,
 one obtains a one-to-one code. This result
 allows us to prove that the average length of our 
 prefix codes ending with a space differs from the minimum
 possible by at most one. In Sections \ref{upper} and \ref{lower},
 we derive upper and lower bounds on the average 
 length  of optimal prefix codes ending with a space  in 
terms of the source entropy and
the cardinality of the code alphabet.}


{
\section{Relations between One-to-One Codes and Prefix Codes Ending with a Space}\label{rel}

Let $S = \{s_1,\ldots,s_n\}$ be the set of
source symbols,  ${\p}=(p_1,\ldots,p_n)$ be a probability distribution on the set $S$
(that is, 
$p_i$ is the 
probability of source symbol $s_i$), and  $\{0,\dots,k-1\}$ be 
the code alphabet. 
We denote by $\{0,\dots,k-1\}^+$ 
the set of all non-empty sequences 
on the code alphabet $\{0,\dots,k-1\}$, $k\geq 2$, and by
$\{0,\dots,k-1\}^+\sqcup$ the set of all non-empty $k$-ary  sequences
that ends with the special symbol $\sqcup$, i.e., the \textit{space} symbol. 
}

{A \textit{prefix-free code} ending with a space
is a one-to-one mapping: 
$$C:S\longmapsto\{0,\dots,k-1\}^+ \cup \{0,\dots,k-1\}^+\sqcup$$
in which no codeword $C(s)$ is a prefix
of another codeword $C(s')$, for any $s,s'\in S$, $s\neq s'$.

 A $k$-ary \textit{one-to-one code} 
 (see \cite{Alon,Blundo,Courtade,Kontoyiannis,Kosut,Szpankowski}
 and the references therein quoted)
 is a bijective mapping 
$D:S\longmapsto  \{0,\dots,k-1\}^+$ from  $S$ to
the set of all non-empty sequences over the alphabet $\{0,\dots,k-1\}$, $k\geq 2$.

The average length of an arbitrary code for the set of source symbols 
$S = \{s_1,\ldots,s_n\}$, with probabilities ${\p}=(p_1,\ldots,p_n)$,
is $\sum_{i=1}^np_i\ell_i$, where $\ell_i$ is the number of alphabet symbols
 in the codeword associated with the source symbol $s_i$.
}

Without loss of generality,
we assume that probability distribution ${\p=(p_1, \ldots ,p_n)}$ is ordered,
that is $p_1\geq\ldots\geq p_n$. Under this assumption,
it is
apparent that the \textit{best} one-to-one code proceeds by
assigning the shortest codeword (e.g., in the binary case, codeword $0$) to the highest probability source symbol $s_1$,
the next shortest codeword $1$  to the 
source symbol $s_2$, the  codeword $00$
to $s_3$,  the  codeword $01$ to $s_4$,  and so on. 

{An equivalent approach for constructing an optimal one-to-one code, which we will use later, proceeds as follows: Let us consider the first $n$ non-empty $k$-ary strings according to the \textit{radix} order \cite{Knuth} (that is, the
$k$-ary strings are ordered by length and, for equal lengths,
ordered according to the lexicographic order). We assign the strings to the symbols 
$s_1,\ldots,s_n$ in $S$ by increasing the string length and, for equal lengths, by inverse order according to the lexicographic order. For example, in the binary case, 
we assign the codeword $1$ to the highest probability source symbol $s_1$, the codeword $0$ to the source symbol $s_2$, the codeword $11$ to $s_3$, the codeword $10$ to $s_4$, and so on.
}
Therefore, one can see that, in the general case of a  $k$-ary code alphabet,  $k\geq 2$, an optimal one-to-one code of
minimal average length assigns a codeword of length $\ell_i$
to the $i$-th symbol $s_i\in S$, 
where $\ell_i$ is given by:
\begin{equation}\label{nonempty}
    \ell_i = \lfloor \log_k ((k-1)\,i + 1)\rfloor.
\end{equation}
{Moreover, 
 the codewords of an optimal $k$-ary one-to-one code can be represented as the nodes of a $k$-ary tree of
maximum depth $h=\lceil\log_k (n-\lceil n/k\rceil)\rceil$, where, for each node $v$, the $k$-ary string (codeword) associated with $v$ is 
obtained by concatenating all the labels in the path from the root of the tree to $v$. }

It is  evident that, if we apply
the above encoding to a \textit{sequence} of source
symbols, the obtained binary sequence is \textit{not} uniquely parsable
in terms of individual codewords.
Let us see how one can recover unique parsability by appending a space $\sqcup$ to judiciously chosen codewords of an {optimal one-to-one code}. 
To gain insight, let us consider the following example. 
{Let $S=\{s_1,s_2,s_3,s_4,s_5,s_6, s_7, s_8,s_9,s_{10}\}$ be the 
set of source symbols, and 
and let us assume that the code alphabet is 
$\{0,1\}$.
Under the standing hypothesis that 
$p_1\geq\ldots\geq p_{10}$, one has 
that the {best} prefix-free code $C$ one can obtain  by 
the procedure of appending a space $\sqcup$ to codewords of the 
optimal one-to-one code for $S$
is the following:
\begin{align*}
    C(s_1) &= 1\sqcup\\
    C(s_2) &= 0\sqcup\\
    C(s_3) &= 11\\
    C(s_4) &= 10\\
    C(s_5) &= 01\sqcup\\
    C(s_6) &= 00\sqcup\\
    C(s_7) &= 011\\
    C(s_8) &= 010\\
    C(s_9) &= 001\\
    C(s_{10}) &= 000.
\end{align*}
Observe that we started from the codewords of the optimal one-to-one code constructed according to the \textit{second} procedure previously described. Moreover, note that the codewords associated with symbols $s_1,s_2,s_5$, and $s_6$ necessitate the space character $\sqcup$ at their end; otherwise, the unique parsability of some encoded sequences of source symbols would not be guaranteed. On the other hand, the codewords associated with symbols $s_3,s_4,s_7,s_8,s_9$, and $s_{10}$ do not necessitate the space character $\sqcup$. Indeed, the codeword set $$\{1\sqcup,0\sqcup,11,10,01\sqcup,00\sqcup,011,010,001,000\}$$  satisfies the prefix-free condition (i.e., no codeword is a prefix of any other); therefore, it guarantees the unique parsability of {any} coded message in terms of individual codewords.
}

The idea of the above example can be
 generalized, {as shown in the following lemma}.

{

\begin{Lemma}\label{lemma:ub_code_wt_space}
    Let $S=\{s_1,\ldots,s_n\}$ be the set of source symbols and 
    ${\p}=(p_1,\ldots , p_n)$,   $p_1\geq\ldots \geq p_n>0$, be a probability
    distribution on $S$.
    Let $\{0,\dots,k-1\}$ be the  $k\geq 2$-ary 
    code alphabet. We can construct  a prefix-free code $C:S\longmapsto\{0,\dots,k-1\}^+ \cup \{0,\dots,k-1\}^+\sqcup$, in $O(n)$, such that its average length $L(C)$ satisfies
    \begin{align}
        \label{eq:ub_space_lemma}
        L(C) &= L_{+} + \sum_{i=1}^{\frac{k^{{h}-1}-1}{k-1}-1} \!\!\!\!p_i  + \sum_{i=\frac{k^{{h}}+k^{{h}-1}-2}{k-1} -\lceil n/k\rceil}^{\frac{k^{{h}}-1}{k-1}-1}
         \!\!\!\!\!\!\!\!\!\!\!\!\!\!p_i \\
        &\leq  L_{+} + \sum_{i=1}^{\lceil n/k\rceil-1} p_i, \label{LC<L+}
    \end{align}
where $L_+$ is the average length of an optimal one-to-one code $D:S\longmapsto  \{0,\dots,k-1\}^+$ {and $h=\lceil\log_k (n-\lceil n/k\rceil)\rceil$}.
\end{Lemma}
\begin{proof}
{
Under the standing hypothesis that the probabilities of the source symbols are ordered from the largest to the smallest, we show how
to construct a prefix-free code---by appending the special character $\sqcup$ to the end of (some) codewords of an optimal one-to-one code for $S$---having the average length upper bounded by (\ref{eq:ub_space_lemma}). 

Among the class of
all the prefix-free codes that one can obtain by appending the character  $\sqcup$ to the end of (some) codewords of an optimal one-to-one code for $S$, we aim to construct the one with the minimum average length. Therefore, we have to ensure that, in the $k$-ary tree representation of the code, the following basic condition holds:
For any pair of nodes $v_i$ and $v_j$, $i<j$, associated with the symbols $s_i$ and $s_j$, the depth of the node $v_j$ is not smaller than the depth of the node $v_i$. In fact, if it were otherwise,
the average length of the code could be improved.}

{Therefore, by recalling that $h=\lceil\log_k (n-\lceil n/k\rceil)\rceil$ is the height of the $k$-ary tree associated with an optimal one-to-one code}, we have that the  prefix-free code 
of  the {minimum} average length that one can obtain by appending the special character
$\sqcup$ to the end of (some) codewords of an optimal one-to-one code for 
$S$ assigns a codeword of length $\ell_i$
to the $i$-th symbol $s_i\in S$, 
where $\ell_i$ is given by:

\begin{equation}
    \label{eq:lengths}
    \ell_i = 
    \begin{cases}
        \lfloor \log_k ((k-1)\,i + 1)\rfloor + 1, &
        \mbox{if $i\leq \frac{k^{{h}-1}-1}{k-1}-1$},\\
        \lfloor \log_k ((k-1)\,i + 1)\rfloor + 1, &\mbox{if } 
        i\geq  \frac{k^{{h}}+k^{{h}-1}-2}{k-1} -\lceil n/k\rceil\\
        &\mbox{ and } i\leq \frac{k^{{h}}-1}{k-1}-1,\\
        \lfloor \log_k ((k-1)\,i+1)\rfloor, &\mbox{otherwise}.
    \end{cases}
\end{equation}
We stress that the obtained prefix-free code is not necessarily a  prefix-free code $C:S\longmapsto\{0,\dots,k-1\}^+ \cup \{0,\dots,k-1\}^+\sqcup$ of minimum average length. Now, we justify the expression  (\ref{eq:lengths}). First, since the probabilities $p_1, \ldots , p_n$
are ordered in non-increasing fashion, the codeword lengths $\ell_i$ of the code are ordered in non-decreasing fashion, that is $\ell_1\leq \ldots  \leq\ell_n$. {Therefore, in the $k$-ary tree representation of the code, it holds the desired basic condition:
For any pair of nodes $v_i$ and $v_j$, $i<j$, associated with the symbols $s_i$ and $s_j$, the depth of the node $v_i$ is smaller than or equal to the depth of the node $v_j$.}

Furthermore,  we need to append the space character only to the $k$-ary strings that
are  the prefix of some others. Therefore, let us consider the first $n$ non-empty $k$-ary strings according to the \textit{radix} order \cite{Knuth}, in which, we recall, the
$k$-ary strings are ordered by length and, for equal lengths,
ordered according to the lexicographic order. We have that the number of strings that are a prefix of some others is exactly $\lceil\frac{n}{k}\rceil-1$. 
One obtains this number by seeing the strings as
corresponding to nodes in a $k$-ary tree with labels $0,\dots,k-1$ on the edges. The number of strings that are a prefix of some others (among the
$n$ strings) is \textit{exactly} equal to the number of internal nodes 
(except the root) in such a tree. 
This number of internal nodes is equal to $\lceil\frac{N-1}{k}\rceil-1$, where $N$ is the total number of nodes that, in our case, is equal to 
$N=n+1$ (i.e., $N$ counts also the root of the tree). 

Moreover, starting from the optimal one-to-one code constructed according
to our second method, that is by assigning $k$-ary strings to the symbols by increasing length and, for equal lengths, by inverse order according to the lexicographic order, one can verify that the $\lceil \frac{n}{k}\rceil - 1$ internal nodes are associated with the codewords of the symbols 
$s_i$, for $i$ that goes from $1$ to $\frac{k^{{h}-1}-1}{k-1}-1$, and from $\frac{k^{{h}}+k^{{h}-1}-2}{k-1} -\lceil n/k\rceil$ to $\frac{k^{{h}}}{k-1}-2$. 
 
 In fact, since the height of the $k$-ary tree is ${h}=\lceil\log_k (n-\lceil n/k\rceil)\rceil$ and since all the levels of the tree, except the last two, are full, we need to append the space to all symbols from $1$ to $\frac{k^{{h}-1}-1}{k-1}-1$. While on the second-to-last level, we have to append the space only to the remaining internal nodes associated with the symbols $s_i$, where $i$  goes from $\frac{k^{{h}}+k^{{h}-1}-2}{k-1} -\lceil n/k\rceil$ to $\frac{k^{{h}}-1}{k-1}-1$. {Those remaining
 nodes are exactly, among all the nodes in the second-to-last level,  the ones associated with the symbols that have smaller probabilities}. Thus, we obtain (\ref{eq:lengths}).

Summarizing, 
we can construct a prefix-free code $C:S\longmapsto\{0,\dots,k-1\}^+ \cup \{0,\dots,k-1\}^+\sqcup$,  in $O(n)$ time, with lengths defined as in (\ref{eq:lengths}), starting from an optimal one-to-one code. Thus:
\begin{align*}
    L(C) =& \sum_{i=1}^n p_i\ell_i \\
    =& \sum_{i=1}^n p_i \lfloor \log_k ((k-1)\,i + 1)\rfloor + \sum_{i=1}^{\frac{k^{{h}-1}-1}{k-1}-1} 
    \!\!\!\!p_i+\sum_{i=\frac{k^{{h}}+k^{{h}-1}-2}{k-1} -\lceil n/k\rceil}^{\frac{k^{{h}}-1}{k-1}-1}
    \!\!\!\!\!\!\!\!\!\!\!\!\!\!p_i \\
    =& L_{+} + \sum_{i=1}^{\frac{k^{{h}-1}-1}{k-1}-1} 
     \!\!\!\!p_i  + \sum_{i=\frac{k^{{h}}+k^{{h}-1}-2}{k-1} -\lceil n/k\rceil}^{\frac{k^{{h}}-1}{k-1}-1}
       \!\!\!\!\!\!\!\!\!\!\!\!\!\!p_i \\
    \leq& L_{+}+\sum_{i=1}^{\lceil n/k\rceil-1} \!\!\!\!\! p_i 
    \quad\mbox{(since we are adding $\lceil n/k\rceil-1$ $p_i$'s, and the $p_i$'s are ordered).}
\end{align*}
\end{proof}
Note that, from Lemma \ref{lemma:ub_code_wt_space}, we obtain that the average length of any \textit{optimal} (i.e., of minimum average length) prefix-free code ending with a space is upper bounded by the formula (\ref{eq:ub_space_lemma}). Furthermore, we have an upper bound on the average length of the optimal prefix-free codes ending with a space in terms of the average length of optimal one-to-one codes.
\medskip

We can also derive a \textit{lower} bound on the average length of optimal prefix-free codes ending with a space in terms of the average length of optimal one-to-one codes. For such a purpose, we need two intermediate results. We first recall that, given a $k$-ary code $C$, its codewords can be represented as nodes in a $k$-ary tree with labels $0,\dots,k-1$ on the edges. Indeed, for each node $v$, the $k$-ary string (codeword) associated with $v$ can be
obtained by concatenating all the labels in the path from the root of the tree to $v$. 
We  also recall that, in  prefix-free codes, the codewords correspond to 
the node leaves of the associated tree, while in one-to-one codes,
the codewords may correspond also  to the internal
nodes of the associated tree.

\begin{Lemma}\label{lemma:internal_nodes}
     Let $S=\{s_1,\ldots,s_n\}$ be the set of source symbols, and let
    ${\p}=(p_1,\ldots , p_n)$, $p_1\geq\ldots \geq p_n>0$, be a probability
    distribution on $S$. 
    There \textit{exists} an \emph{optimal prefix-free} code ending with a space $C:S\longmapsto\{0,\dots,k-1\}^+ \cup \{0,\dots,k-1\}^+\sqcup$ such that the following property holds:
    For any internal node $v$ (except the root) of the tree representation of $C$, 
    if we denote by $w$ the $k$-ary string associated with the node $v$, then the
    string $w\sqcup$ belongs to the codeword set of $C$.
\end{Lemma}
\begin{proof}
    Let $C$ be an arbitrary optimal prefix-free code ending with a space.
   Let us assume that,
    in the tree representation of $C$, there exists an internal node $v$ 
    whose associated string $w$ is such that 
    $w\sqcup$ \textit{does not} belong to the codeword set of $C$.
    Since $v$ is an internal node, there is at least a leaf $x$, which is a descendant of $v$,  whose associated string is the codeword of some symbol $s_j$. 
    We modify the encoding, by assigning the codeword $w\sqcup$ to the symbol $s_j$. The new encoding  is still prefix-free, and its average length can only decrease since the 
    length of the newly assigned codeword to $s_j$ cannot be greater than the previous one. 
    We can repeat the argument for all internal nodes that do not satisfy the
    property stated in the lemma to complete the proof.
\end{proof}

\begin{Lemma}\label{lemma:removing_space_from_prefix}
    Let $C:S\longmapsto\{0,\dots,k-1\}^+ \cup \{0,\dots,k-1\}^+\sqcup$ be an arbitrary prefix-free code, then the code $D:S\longmapsto\{0,\dots,k-1\}^+$ 
    one obtains from $C$ by removing the space $\sqcup$ from each 
    codeword of $C$ is a \emph{one-to-one }code.
\end{Lemma}
\begin{proof}
    The proof is straightforward. Since $C$ is prefix-free,  it holds that, for any pair $s_i,s_j\in S$, with $s_i\neq s_j$, the codeword $C(s_i)$ is not a prefix of $C(s_j)$ and vice versa. Therefore, since $D$ is obtained from $C$ by removing the space, we have four cases:
    
    \begin{enumerate}
        \item $C(s_i)=D(s_i)$ and $C(s_j)=D(s_j)$: then $D(s_i) \neq D(s_j)$ since $C(s_i)\neq C(s_j)$;
        \item $C(s_i)=D(s_i)\sqcup$ and $C(s_j)=D(s_j)\sqcup$: then $D(s_i) \neq D(s_j)$ since $C(s_i)$ is not a prefix of $C(s_j)$ and vice versa;
        \item $C(s_i)=D(s_i)\sqcup$ and $C(s_j)=D(s_j)$: then $D(s_i) \neq D(s_j)$ since $C(s_j)$ is not a prefix of $C(s_i)$;
        \item $C(s_i)=D(s_i)$ and $C(s_j)=D(s_j)\sqcup$: then $D(s_i) \neq D(s_j)$ since $C(s_i)$ is not a prefix of $C(s_j)$.
    \end{enumerate}
    Therefore, for any pair $s_i,s_j\in S$, with $s_i\neq s_j$, $D(s_i) \neq D(s_j)$, and $D$ is a one-to-one code.
\end{proof}

We can now derive a lower bound on the average length of optimal prefix-free codes with space in terms of the average length of optimal one-to-one codes.

\begin{Lemma}\label{lemma:lb_space}
    Let $S=\{s_1,\ldots,s_n\}$ be the set of source symbols, and let
    ${\p}=(p_1,\ldots , p_n)$, $p_1\geq\ldots \geq p_n>0$, be a probability
    distribution on $S$, then the average of an optimal prefix-free code $C:S\longmapsto\{0,\dots,k-1\}^+ \cup \{0,\dots,k-1\}^+\sqcup$ satisfies
    \begin{equation}\label{eq:low}
        L(C) \geq L_{+} + \sum_{i=1}^{\lceil n/k \rceil-1} p_{n-i+1},
    \end{equation}
where $L_{+}$ is the average length of an optimal $k$-ary one-to-one code on $S$.
\end{Lemma}
\begin{proof}
    From Lemma \ref{lemma:internal_nodes}, we know that there exists an optimal prefix-free code $C$ with a space in which exactly $\lceil \frac{n}{k}\rceil-1$ codewords contain the space character at the end. Let $A\subset \{1,\dots,n\}$ be the set of indices associated with the symbols whose codeword contains the space. Moreover, from Lemma \ref{lemma:removing_space_from_prefix}, we know that the code $D$ obtained by removing the space from $C$ is a one-to-one code. Putting it all together, we obtain that
    \begin{align}\label{eq:L(D)}
        L(D) = L(C)-\sum_{i\in A} p_i.
    \end{align}
    From (\ref{eq:L(D)}), we have that
    \begin{align*}
        L(C) &= L(D) + \sum_{i\in A} p_i\\
        &\geq L_{+} + \sum_{i\in A} p_i\qquad\mbox{(since $D$ is a one-to-one code)}\\
        &\geq  L_{+} + \sum_{i=1}^{\lceil n/k \rceil-1} \!\!\! p_{n-i+1} \quad\mbox{(since $A$ contains $\lceil\frac{n}{k}\rceil-1$ elements).}
    \end{align*}
\end{proof}

We notice that the difference between the expression (\ref{eq:ub_space_lemma}) 
 and the lower bound (\ref{eq:low}) is, because of (\ref{LC<L+}),
less than 
\begin{equation}\label{diff}
\sum_{i=1}^{\lceil n/k\rceil-1} p_i-\sum_{i=1}^{\lceil n/k \rceil-1} p_{n-i+1}<1;
\end{equation}
therefore, the prefix-free codes ending with a space that we 
construct in Lemma \ref{lemma:ub_code_wt_space} have an average length that 
differs from the minimum possible by at most one.
Moreover, since 
both the upper bound (\ref{LC<L+}) and the lower bound (\ref{eq:low}) are 
easily computable,
we can  determine the 
average length of an \textit{optimal} prefix-free code $C:S\longmapsto\{0,\dots,k-1\}^+ \cup \{0,\dots,k-1\}^+\sqcup$ with a tolerance 
at most of one. One can also see that the 
left-hand side of (\ref{diff}) is, often, much smaller than one.

In the following sections, we will focus on providing upper and lower bounds on the average length $L_{+}$ of $k$-ary optimal one-to-one codes in terms of the $k$-ary Shannon entropy $H_k({\p})=-\sum_{i=1}^np_i\log_kp_i$ of the source distribution
${\p}$.
Because of  Lemma \ref{lemma:ub_code_wt_space} and Lemma \ref{lemma:lb_space},
this will give us the 
corresponding upper and lower bounds on the average length of optimal prefix-free codes ending with a space.
}

\section{Lower Bounds on the Average Length}\label{lower}

{ In this section, we provide lower bounds on the average length of the optimal one-to-one code and, subsequently, thanks to Lemma \ref{lemma:lb_space}, on the average length of the optimal prefix-free code with a space.
}
For technical reasons, it will be convenient  
to consider one-to-one codes that make use of the empty word $\epsilon$, 
that is
one-to-one mappings
 $D_{\epsilon}: S\longmapsto \{0,1, \ldots , k-1\}^+ \cup \{\epsilon\}$. It is easy to see (cf. (\ref{nonempty}))
 that 
 the optimal one-to-one code that makes use of the empty word
assigns  to the $i$-th symbol $s_i\in S$ 
a codeword of length $\ell_i$ given by:
\begin{equation}
    \ell_i = \lfloor \log_k ((k-1)i)\rfloor.
\end{equation}
where $k$ is the cardinality of the code alphabet.

Thus, by denoting by $L_{+}$  the average length of the  optimal one-to-one code that \textit{does} not make use of the empty word and with $L_{\epsilon}$
the average length of the  optimal one-to-one code that \textit{does}  use  it, we obtain the following relation:
\begin{equation}
    \label{eq:relation_with_eps}
    L_{+} = L_{\epsilon} + \sum_{i=1}^{\lfloor \log_k \lceil \frac{n-1}{k}\rceil\rfloor} \!\!\!\!\!\! p_{\frac{k^i-1}{k-1}}.
\end{equation}

\medskip

Our first result is a generalization of the lower bound to
the average length of the optimal one-to-one codes presented in \cite{Blundo},
from the binary case to the general case of $k$-ary alphabets, $k\geq 2$. 
Our proof  technique differs from that of \cite{Blundo}  since we are dealing
with a set of source symbols of \textit{bounded} cardinality (in \cite{Blundo}, the authors
considered the case of a numerable set of source symbols).
 
\begin{Lemma}
    \label{lb_one_to_one}
    Let $S=\{s_1,\ldots,s_n\}$ be the set of source symbols and 
    ${\p}=(p_1,\ldots , p_n)$ be a probability
    distribution on $S$, with $p_1\geq\ldots \geq p_n$.
    The average length $L_\epsilon$ of the optimal  one-to-one code $D:\{s_1,\dots,s_n\} \to \{0,\dots,k-1\}^+ \cup \{\epsilon\}$ satisfies
    \begin{equation}
        \label{eq:lb_one_to_one}
        \begin{aligned}         
       L_\epsilon >& H_k({\p}) -(H_k({\p})+\log_k(k-1))\log_k\left(1 + \frac{1}{H_k({\p}) + \log_k(k-1)}\right)\nonumber\\ 
        &-\log_k(H_k({\p})+\log_k(k-1)+1),
        \end{aligned}   
    \end{equation}
where
    $H_k({\p})=-\sum_{i=1}^np_i\log_kp_i$.
\end{Lemma}
\begin{proof}
    The proof is an adaptation of Alon {et al.}'s proof  \cite{Alon} from the binary case to the $k\geq 2$-ary case.
    
   We recall that the 
   optimal one-to-one code  
   (i.e., whose average length  achieves 
    the minimum $L_{\epsilon}$)
    has codeword lengths $\ell_i$  given by:
    \begin{equation}\label{l}
       \ell_i = \lfloor \log_k((k-1)i) \rfloor.
    \end{equation}
    For each $j\in \{0,\dots,\lfloor\log_kn\rfloor\}$, let us define the quantities $q_j$ as
    $$q_j = \sum_{i=\frac{k^j -1}{k-1} + 1}^{\frac{k^{j+1}-1}{k-1}} \!\!\!\!\! p_i.$$ 
    It holds that $\sum_{j=0}^{\lfloor\log_kn\rfloor}q_j=1.$
    Let ${Y}$ be a random variable that takes  values in 
    $\{0,\dots,\lfloor\log_kn\rfloor\}$ according to the probability distribution 
    ${\q}=(q_0,\dots,q_{\lfloor\log_kn\rfloor})$, that is 
    $$\forall j\in \{0,\dots,\lfloor\log_kn\rfloor\} \quad  \Pr\{{Y}=j\}=q_j.$$
    From (\ref{l}), we have 
        \begin{align} 
            L_{\epsilon} =& \sum_{i=1}^n \lfloor \log_k((k-1)i) \rfloor p_i\nonumber\\
            =& \sum_{j=0}^{\lfloor\log_kn\rfloor} \sum_{i=\frac{k^j -1}{k-1} + 1}^{\frac{k^{j+1}-1}{k-1}} \lfloor \log_k((k-1)i) \rfloor p_i\nonumber\\
            =&\sum_{j=0}^{\lfloor\log_kn\rfloor} j q_j = \mathbb{E}[{Y}]\label{LeqQ}.
        \end{align}
  
    By applying the entropy grouping rule (\cite{CT}, Ex. 2.27) to the distribution ${\p}$,
    we obtain
    \begin{align}
        \label{eq:ub_respect_to_q}
        H_2({\p}) &= H_2({\q}) + \sum_{j=0}^{\lfloor \log_kn \rfloor} q_j H_2\left(\frac{p_{\frac{k^j-1}{k-1} + 1}}{q_j},\dots, \frac{p_{\frac{k^{j+1}-1}{k-1}}}{q_j}\right)\nonumber\\
        &\leq H_2({\q}) + \sum_{j=0}^{\lfloor \log_kn \rfloor} q_j \log_2k^j\nonumber\hspace{1cm} \mbox{(since $H_2\left(\frac{p_{\frac{k^j-1}{k-1} + 1}}{q_j},\dots, \frac{p_{\frac{k^{j+1}-1}{k-1}}}{q_j}\right) \leq \log_2k^j$)}\\
        &= H_2({\q}) + \sum_{j=0}^{\lfloor \log_kn \rfloor} j q_j \log_2k\nonumber\\
        &= H_2({\q}) + \mathbb{E}[{Y}]\log_2k.
    \end{align}
   We now derive 
   an upper bound to $H_2({Y})=H_2(q)$ in terms of the expected value $\mathbb{E}[{Y}]$.
    
    To this end, let us consider an auxiliary random variable ${Y}'$ with the same distribution of ${Y}$, but with values ranging from $1$ to $\lfloor\log_k(n)\rfloor + 1$ (instead of from $0$ to $\lfloor\log_k(n)\rfloor$). 
    It is easy to verify that $\mu = \mathbb{E}[{Y}']=\mathbb{E}[{Y}]+1$.

    Let $\alpha$ be a positive number, whose value will be chosen later. We obtain
    that
    \begin{equation*}
        \begin{aligned}
            H_k({Y}) - \alpha\mu &= \sum_{i=1}^{\lfloor\log_k(n)\rfloor + 1} q_{i-1} \log_k\frac{1}{q_{i-1}} - \alpha\sum_{j=1}^{\lfloor\log_k(n)\rfloor + 1}j q_{j-1}\\
            &= \sum_{i=1}^{\lfloor\log_k(n)\rfloor + 1} q_{i-1}\log_k\frac{1}{q_{i-1}} + \sum_{j=1}^{\lfloor\log_k(n)\rfloor + 1} (-\alpha j) q_{j-1}\\
            &=  \sum_{i=1}^{\lfloor\log_k(n)\rfloor + 1} q_{i-1} \log_k\frac{1}{q_{i-1}} + \sum_{j=1}^{\lfloor\log_k(n)\rfloor + 1} q_{j-1} \log_k(k^{-\alpha j})\\
            &= \sum_{i=1}^{\lfloor\log_k(n)\rfloor + 1} q_{i-1} \log_k \frac{k^{-\alpha i}}{q_{i-1}} \\
            &\leq \log_k \sum_{i=1}^{\lfloor\log_k(n)\rfloor + 1} k^{-\alpha i}
            \qquad\qquad \mbox{( by Jensen's inequality)}\\
            &= \log_k \left[\left(\frac{1}{k^\alpha}\right) \left(\frac{1-k^{-\alpha (\lfloor\log_k(n)\rfloor+1}}{1-k^{-\alpha}}\right)\right]\\
            &\leq \log_k\left(\frac{1-k^{-\alpha (\log_k(n)+1)}}{k^\alpha-1}\right)\\
            &= \log_k\left(\frac{1-(k n)^{-\alpha}}{k^\alpha-1}\right).
        \end{aligned}
    \end{equation*}
  By substituting  $\log_k \frac{\mu}{\mu-1}$ with 
  $\alpha$ in the obtained inequality
  $$H_k({Y})\leq \alpha\mu+\log_k\left(\frac{1-(k n)^{-\alpha}}{k^\alpha-1}\right),$$
we obtain
    \begin{equation}
        \label{eq:ub_of_lemma_1}
        \begin{aligned}
            H_k({Y}) \leq \mu \log_k \frac{\mu}{\mu-1}+\log_k(\mu-1)+\log_k\left(1-\left(\frac{1}{k n}\right)^{\log_k \frac{\mu}{\mu-1}}\right).
        \end{aligned}
    \end{equation}
   Since $\left(\frac{1}{k n}\right)^{\log_k \frac{\mu}{\mu-1}}$ is decreasing in $\mu$,
   and because $\mu = \mathbb{E}[{Y}]+1>1 $, we  obtain:
    \begin{equation}
        \label{eq:upper_bound_to_Hk}
        \begin{aligned}
            H_k({Y}) <\mathbb{E}[{Y}] \log_k\left(1+\frac{1}{\mathbb{E}[{Y}]}\right) + \log_k(\mathbb{E}[{Y}]+1).
        \end{aligned}
    \end{equation}
    By applying (\ref{eq:upper_bound_to_Hk}) to (\ref{eq:ub_respect_to_q}) and since
     $H_k({Y}) = \frac{H_2({Y})}{\log_2 k}$, we obtain
    \begin{align}                   \label{eq:expected_value_inequality}    
            H_2({\p}) <& \mathbb{E}[{Y}]\log_2 k + \mathbb{E}[{Y}]\log_2\left(1+\frac{1}{E[{Y}]}\right) + \log_2(E[{Y}]+1).
    \end{align}
    
   From (\ref{LeqQ}), we have that $L_{\epsilon} = \mathbb{E}[{Y}]$;  moreover,  from the 
    inequality (\ref{eq:lemma4}) of  Lemma \ref{Wyner_gen} (proven in the 
     next Section \ref{upper}),  we know that 
    \begin{equation}
        \label{eq:upper_bound}
        L_{\epsilon} \leq H_k({\p})+\log_k(k-1).
    \end{equation}
    Hence, since the function $f(z)=z\log_k\left(1+\frac{1}{z}\right)$ is increasing in $z$, we can apply (\ref{eq:upper_bound}) to  upper-bound the term
    $$\mathbb{E}[{Y}]\log_2\left(1+\frac{1}{E[{Y}]}\right),$$
    to  obtain
    the following inequality:
    \begin{equation}\label{w}
        \begin{aligned}
            H_2({\p}) <& L_{\epsilon} \log_2 k + (H_k({\p}) +\log_k(k-1))\log_2\left(1+\frac{1}{H_k({\p}) +\log_k(k-1)}\right)\\
            &+ \log_2(H_k({\p}) +\log_k(k-1) + 1).
        \end{aligned}
    \end{equation}
    Rewriting (\ref{w}), we finally obtain
    \begin{equation*}
        \begin{aligned}         
        L_{\epsilon} >& H_k({\p}) -(H_k({\p})+\log_k(k-1)) \log_k\left(1 + \frac{1}{H_k({\p}) + \log_k(k-1)}\right)\\ 
        &-\log_k(H_k({\p})+\log_k(k-1)+1), 
        \end{aligned}   
    \end{equation*}
and that concludes our proof.
\end{proof}

{By bringing into play the size of the largest mass in addition to the
entropy,} Lemma \ref{lb_one_to_one} can be improved, as shown in the following result.
\begin{Lemma}
    \label{lb_one_to_one_2}
    Let $S=\{s_1,\ldots,s_n\}$ be the set of source symbols and 
    ${\p}=(p_1,\ldots , p_n)$,   $p_1\geq\ldots \geq p_n$, be a probability
    distribution on $S$.
     The average length $L_\epsilon$ of the optimal  one-to-one code $D:\{s_1,\dots,s_n\} \to \{0,\dots,k-1\}^+ \cup \{\epsilon\}$ has the following lower bounds:
    \medskip
    
    \textbf{1.} If $0<p_1\leq 0.5$, 
        
        \begin{align}\label{eq:lb_one_to_one_2_1}        
        L_\epsilon \geq& H_k({\p}) -(H_k({\p})-p_1\log_k \frac{1}{p_1} +(1-p_1) \log_k(k-1))\nonumber\\
        &\log_k\left(1 + \frac{1}{H_k({\p}) -p_1\log_k \frac{1}{p_1} +(1-p_1) \log_k(k-1)}\right)\nonumber\\ 
        &-\log_k(H_k({\p})-p_1\log_k \frac{1}{p_1} +(1-p_1) \log_k(k-1)+1)\nonumber\\&
        -\log_k\left(1-\left(\frac{1}{k n}\right)^{\log_k\left(1+\frac{1}{1-p_1}\right)}\right),
        \end{align}  
   2. \textbf{if} $0.5<p_1\leq1$ 
    \begin{align}
        \label{eq:lb_one_to_one_2_2}      
        L_\epsilon \geq& H_k({\p}) -(H_k({\p})-\mathcal{H}_k(p_1) +(1-p_1) (1+\log_k(k-1))) \nonumber\\
        &\log_k\left(1 + \frac{1}{H_k({\p}) -\mathcal{H}_k(p_1) +(1-p_1) (1+\log_k(k-1))}\right)\nonumber\\ 
        &-\log_k(H_k({\p})-\mathcal{H}_k(p_1) +(1-p_1) (1+\log_k(k-1))+1)\nonumber\\
        &-\log_k\left(1-\left(\frac{1}{k n}\right)^{\log_k\left(1+\frac{1}{1-p_1}\right)}\right), 
    \end{align}
where $\mathcal{H}_k(p_1)=
    -p_1\log_k p_1 -(1-p_1)\log_k(1-p_1)$.
\end{Lemma}
\begin{proof}
    The proof is the same as the proof of Lemma \ref{lb_one_to_one}.
    However, we change two steps in the demonstration. 
    
    First, since 
    $$
    \left(\frac{1}{kn}\right)^{\log_k \frac{\mu}{\mu-1}} = \left(\frac{1}{kn}\right)^{\log_k \frac{\mathbb{E}[{Y}] + 1}{\mathbb{E}[{Y}]}} = \left(\frac{1}{kn}\right)^{\log_k \left(1 + \frac{1}{\mathbb{E}[{Y}]}\right)}
    $$
is decreasing in $\mu$ and $\mathbb{E}[{Y}] = L_{\epsilon} = 0 p_1 + 1 p_2 + \dots \geq 1-p_1$, we have
    \begin{equation}
        \label{eq:ub_depending_on_p_1}
        \log_k\left(1-\left(\frac{1}{k n}\right)^{\log_k \frac{\mu}{\mu-1}}\right) \leq \log_k\left(1-\left(\frac{1}{k n}\right)^{\log_k \left(1+\frac{1}{1-p_1}\right)}\right).
    \end{equation}
    Hence, by applying (\ref{eq:ub_depending_on_p_1}) to the right-hand side of  (\ref{eq:ub_of_lemma_1}), we obtain 
    \begin{equation}
        \label{eq:ub_of_lemma_2}
        H_k({Y}) \leq \mathbb{E}[{Y}] \log_k\left(1+\frac{1}{\mathbb{E}[{Y}]}\right) + \log_k(\mathbb{E}[{Y}]+1) + \log_k\left(1-\left(\frac{1}{k n}\right)^{\log_k \left(1+\frac{1}{1-p_1}\right)}\right).
    \end{equation}
    Now, by applying (\ref{eq:ub_of_lemma_2}) (instead of (\ref{eq:upper_bound_to_Hk})) to (\ref{eq:ub_respect_to_q}) and since
     $H_k({Y}) = \frac{H_2({Y})}{\log_2 k}$, we obtain
    \begin{align}                           \label{eq:expected_value_inequality_lemma_2}    
            H_2({\p}) \leq& \mathbb{E}[{Y}]\log_2 k + \mathbb{E}[{Y}]\log_2\left(1+\frac{1}{E[{Y}]}\right) + \log_2(E[{Y}]+1)\nonumber\\
            &+\log_2\left(1-\left(\frac{1}{k n}\right)^{\log_k \left(1+\frac{1}{1-p_1}\right)}\right).
    \end{align}
    
    Here, instead of applying the upper bound: 
    \begin{equation*}
        L_{\epsilon} \leq H_k({\p})+\log_k(k-1)
    \end{equation*}
of Lemma \ref{Wyner_gen} to the right-hand side of (\ref{eq:expected_value_inequality_lemma_2}), we apply the improved version:
    \begin{equation*}
        L_{\epsilon} \leq 
        \begin{cases}
        H_k({\p}) -p_1\log_k \frac{1}{p_1} + (1-p_1) \log_k(k-1) \text{ if }\  0<p_1\leq 0.5,\\
        H_k({\p}) -\mathcal{H}_k(p_1) + (1-p_1) \log_k 2 (k-1)\text{ if } \ 0.5<p_1\leq 1,
        \end{cases} 
    \end{equation*}
proven in  Lemma \ref{Blundo_gen} of the Section \ref{upper}. Then, we simply need to rewrite the inequality, concluding the proof.
\end{proof}

Thanks to {Lemma \ref{lemma:lb_space}} and the formula (\ref{eq:relation_with_eps}), the above lower bounds on $L_\epsilon$ can be applied 
to derive our main results for prefix-free codes with a space,  as shown in the following theorems.
\begin{Theorem}
    The average length of an optimal prefix-free code with space $C:S\longmapsto\{0,\dots,k-1\}^+ \cup \{0,\dots,k-1\}^+\sqcup$
     satisfies
    \begin{align} 
        \label{eq:th_1}
        L(C) >& H_k({\p}) -(H_k({\p})+\log_k(k-1)) \log_k\left(1 + \frac{1}{H_k({\p}) + \log_k(k-1)}\right)\nonumber\\ 
        &-\log_k(H_k({\p})+\log_k(k-1)+1) + \sum_{i=1}^{\lceil \frac{n}{k}\rceil-1} {p_{n-i+1}} + \sum_{i=1}^{\lfloor \log_k \lceil \frac{n-1}{k}\rceil\rfloor} p_{\frac{k^i-1}{k-1}}.
    \end{align}
\end{Theorem}
\begin{proof}
    From Lemma {\ref{lemma:lb_space}} and the formula (\ref{eq:relation_with_eps}), we have
    \begin{equation}
        \label{eq:length_of_optimal_prefix_free}
        L(C) {\geq} L_{\epsilon} + \sum_{i=1}^{\lceil\frac{n}{k}\rceil-1} {p_{n-i+1}} + \sum_{i=1}^{\lfloor \log_k \lceil \frac{n-1}{k}\rceil\rfloor} p_{\frac{k^i-1}{k-1}}.
    \end{equation}
    By applying the lower bound (\ref{eq:lb_one_to_one}) of Lemma \ref{lb_one_to_one} to (\ref{eq:length_of_optimal_prefix_free}), we obtain (\ref{eq:th_1}).
\end{proof}

Analogously, by exploiting
(the possible) knowledge of the maximum source symbol probability
value, we have
the following result.

\begin{Theorem}
    The average length of the optimal prefix-free code with space $C:S\longmapsto\{0,\dots,k-1\}^+ \cup \{0,\dots,k-1\}^+\sqcup$ 
    has the following lower bounds:
    
   \textbf{1.}  If $0<p_1\leq 0.5$:
        \begin{align}
    \label{eq:th_2_1}
    L(C) \geq& H_k({\p}) -(H_k({\p})-p_1\log_k \frac{1}{p_1} +(1-p_1) \log_k(k-1)) \nonumber\\
        &\log_k\left(1 + \frac{1}{H_k({\p}) -p_1\log_k \frac{1}{p_1} +(1-p_1) \log_k(k-1)}\right)\nonumber\\ 
        &-\log_k(H_k({\p})-p_1\log_k \frac{1}{p_1} +(1-p_1) \log_k(k-1)+1)\nonumber\\
        &-\log_k\left(1-\left(\frac{1}{k n}\right)^{\log_k\left(1+\frac{1}{1-p_1}\right)}\right)+\sum_{i=1}^{\lceil \frac{n}{k}\rceil-1} {p_{n-i+1}} + \sum_{i=1}^{\lfloor \log_k \lceil \frac{n-1}{k}\rceil\rfloor} p_{\frac{k^i-1}{k-1}}.
\end{align}

\textbf{2.} If $0.5<p_1\leq 1$:

\begin{align}
    \label{eq:th_2_2}
    L(C) \geq& H_k({\p}) -(H_k({\p})-\mathcal{H}_k(p_1) +(1-p_1) (1+\log_k(k-1))) \nonumber\\
        &\log_k\left(1 + \frac{1}{H_k({\p}) -\mathcal{H}_k(p_1) +(1-p_1) (1+\log_k(k-1))}\right)\nonumber\\ 
        &-\log_k(H_k({\p})-\mathcal{H}_k(p_1) +(1-p_1) (1+\log_k(k-1))+1)\nonumber\\
        &-\log_k\left(1-\left(\frac{1}{k n}\right)^{\log_k\left(1+\frac{1}{1-p_1}\right)}\right)+\sum_{i=1}^{\lceil \frac{n}{k}\rceil-1} {p_{n-i+1}} + \sum_{i=1}^{\lfloor \log_k \lceil \frac{n-1}{k}\rceil\rfloor} p_{\frac{k^i-1}{k-1}}.
\end{align}
\end{Theorem}
\begin{proof}
    From Lemma \ref{lemma:lb_space} and the formula (\ref{eq:relation_with_eps}), we have
    \begin{equation}
        \label{eq:length_of_optimal_prefix_free_1}
        L(C) {\geq} L_{\epsilon} + \sum_{i=1}^{\lceil\frac{n}{k}\rceil-1} {p_{n-i+1}} + \sum_{i=1}^{\lfloor \log_k \lceil \frac{n-1}{k}\rceil\rfloor} p_{\frac{k^i-1}{k-1}}.
    \end{equation}
    By applying the lower bounds (\ref{eq:lb_one_to_one_2_1})-(\ref{eq:lb_one_to_one_2_2}) of Lemma \ref{lb_one_to_one_2} to (\ref{eq:length_of_optimal_prefix_free_1}), we obtain (\ref{eq:th_2_1}) or (\ref{eq:th_2_2}) according to the value of the maximum source symbol probability.
\end{proof}

\section{Upper Bounds on the Average Length}\label{upper}

In this section, we will first derive
\textit{upper} bounds on the average length of  optimal one-to-one codes.
Successively, we will provide corresponding upper bounds
on the average length of optimal prefix-free codes ending with a space.

We start by extending the result obtained in \cite{Wyner} from the binary case to the $k$-ary case,  $k \geq 2$.
\begin{Lemma}
    \label{Wyner_gen} 
     Let $S=\{s_1,\ldots,s_n\}$ be the set of source symbols and 
    ${\p}=(p_1,\ldots , p_n)$,   $p_1\geq\ldots \geq p_n$, be a probability
    distribution on $S$.
The average length  $L_\epsilon$ of the optimal  one-to-one code $D:\{s_1,\dots,s_n\} \to \{0,\dots,k-1\}^+ \cup \{\epsilon\}$ satisfies
    \begin{equation}\label{eq:lemma4}
        L_{\epsilon} \leq H_k({\p})+\log_k(k-1).
    \end{equation}
\end{Lemma}
\begin{proof}
   Under the standing hypothesis that $p_1\geq\ldots \geq p_n$, it holds that
    \begin{equation}
        \label{eq:non_increasing_prop}
     \forall  i=1,\ldots,n \qquad  p_i \leq \frac{1}{i}.
    \end{equation}
We recall that  the length of the $i$-th codeword of the optimal one-to-one code $D$ is equal to
    \begin{equation}
        \label{eq:length_of_ith_codeword}
        \ell_i=\lfloor\log_k((k-1) i)\rfloor.
    \end{equation}
    Therefore, from (\ref{eq:non_increasing_prop}), we can 
    upper bound each length $\ell_i$ as 
    \begin{equation}
        \label{eq:ub_on_length_of_ith_codeword}
        \ell_i = \lfloor\log_k((k-1) i)\rfloor \leq \log_k((k-1) i) \leq \log_k(k-1) + \log_k \frac{1}{p_i}.
    \end{equation}
    Hence, by applying (\ref{eq:ub_on_length_of_ith_codeword}) to the average length of $D$, we obtain
    \begin{equation}
        L_{\epsilon} = \sum_{i=1}^n p_i \ell_i \leq \sum_{i=1}^n p_i \left(\log_k(k-1) + \log_k \frac{1}{p_i}\right) = H_k({\p}) + \log_k(k-1).
    \end{equation}
    This concludes our proof.
\end{proof}

By exploiting the knowledge of the
maximum probability value of ${\p}$,
we generalize the upper bound in \cite{Blundo} 
from $k=2$ to  arbitrary $k\geq 2$.

\begin{Lemma}
    \label{Blundo_gen}
       Let $S=\{s_1,\ldots,s_n\}$ be the set of source symbols and 
    ${\p}=(p_1,\ldots , p_n)$,   $p_1\geq\ldots \geq p_n$, be a probability
    distribution on $S$.
The average length  $L_\epsilon$ of the optimal  one-to-one code $D:\{s_1,\dots,s_n\} \to \{0,\dots,k-1\}^+ \cup \{\epsilon\}$ satisfies
    \begin{equation}\label{eq:Blundo_gen}
        L_{\epsilon} \leq 
        \begin{cases}
        H_k({\p}) -p_1\log_k \frac{1}{p_1} + (1-p_1) \log_k(k-1) \text{ if } 0<p_1\leq 0.5,\\
        H_k({\p}) -\mathcal{H}_k(p_1) + (1-p_1) \log_k 2 (k-1)\text{ if } 0.5<p_1\leq 1.
    \end{cases} 
    \end{equation}
\end{Lemma}
\begin{proof}
    Let us prove first that the length of an optimal one-to-one code satisfies the inequality:
    \begin{align}
        \label{upper_bound}
        L_{\epsilon} \leq& 
        \sum_{i=2}^n p_i\log_k(i (k-1)) - 0.5 \sum_{j\geq 2: \frac{k^j-1}{k-1}\leq n} \!\!\!\!\!\! p_{\frac{k^j-1}{k-1}}.
    \end{align}
    Indeed, by recalling that $\ell_1 = \lfloor\log_k(k-1)\rfloor = 0$,
    we can write $L_{\epsilon}$ as follows: 
    \begin{equation*}
        \begin{aligned}
           L_{\epsilon} =& \sum_{i=2}^n p_i\lfloor\log_k(i (k-1)) \rfloor\\
           =& \sum_{j \geq 1: \frac{k^j-1}{k-1}+1 \leq n} \sum_{i=\frac{k^j-1}{k-1}+1}^{\min(\frac{k^{j+1}-1}{k-1}-1, n)} p_i\lfloor\log_k(i (k-1)) \rfloor + \sum_{j\geq 2: \frac{k^j-1}{k-1} \leq n} p_{\frac{k^j -1}{k-1}}\left\lfloor\log_k\left(\frac{k^j-1}{k-1} (k-1)\right) \right\rfloor\\
           =& \sum_{j \geq 1: \frac{k^j-1}{k-1}+1 \leq n} \sum_{i=\frac{k^j-1}{k-1}+1}^{\min(\frac{k^{j+1}-1}{k-1}-1, n)} p_i\lfloor\log_k(i (k-1)) \rfloor +\sum_{j\geq 2: \frac{k^j-1}{k-1}\leq n}  p_{\frac{k^j-1}{k-1}}\log_k(k^j-1)\\
           & -\sum_{j\geq 2: \frac{k^j-1}{k-1}\leq n} p_{\frac{k^j-1}{k-1}} (\log_k(k^j-1)-\lfloor\log_k(k^j-1)\rfloor)\\
           \leq& \sum_{i=2}^n p_i\log_k(i (k-1))-\sum_{j\geq 2: \frac{k^j-1}{k-1}\leq n} p_{\frac{k^j-1}{k-1}} (\log_k(k^j-1)-\lfloor\log_k(k^j-1)\rfloor),
        \end{aligned}
    \end{equation*}
where the last inequality holds since 
    \begin{equation*}
        \sum_{j \geq 1: \frac{k^j-1}{k-1}+1 \leq n} \sum_{i=\frac{k^j-1}{k-1}+1}^{\min(\frac{k^{j+1}-1}{k-1}-1, n)} p_i\lfloor\log_k(i (k-1)) \rfloor \leq \sum_{j \geq 1: \frac{k^j-1}{k-1}+1 \leq n} \sum_{i=\frac{k^j-1}{k-1}+1}^{\min(\frac{k^{j+1}-1}{k-1}-1, n)} p_i\log_k(i (k-1)).
    \end{equation*}
    We note that the function $f(j)=\log_k(k^j-1)-\lfloor\log_k(k^j-1)\rfloor$ is  
    increasing  in $j$. Therefore, it reaches the minimum at $j=2$,
    where it takes the value 
   $$\log_k(k^2-1)-\lfloor\log_k(k^2-1)\rfloor = 1+\log_k\left(1-\frac{1}{k^2}\right) > 0.5,
   $$ 
   for any $k\geq 2$. Thus, (\ref{upper_bound}) holds as we claimed.

    \smallskip
    
    Let us now show that
    \begin{equation}
        \label{eq:first}
        L_{\epsilon} \leq H_k({\p})-p_1\log_k\frac{1}{p_1} +(1-p_1) \log_k(k-1)-0.5 \sum_{j\geq 2: \frac{k^j-1}{k-1}\leq n} p_{\frac{k^j-1}{k-1}}.
    \end{equation}
    Since the distribution
    ${\p}$ is ordered in a non-increasing fashion,   from  (\ref{eq:non_increasing_prop}) and (\ref{upper_bound}), we have
    \begin{equation*}
        \begin{aligned}
            L_{\epsilon} &\leq \sum_{i=2}^n p_i\log_k(i (k-1)) - 0.5 \sum_{j\geq 2: \frac{k^j-1}{k-1}\leq n} p_{\frac{k^j-1}{k-1}}\\
            &\leq \sum_{i=2}^n p_i\log_k\frac{1}{p_i} (k-1)- 0.5 \sum_{j\geq 2: \frac{k^j-1}{k-1}\leq n} p_{\frac{k^j-1}{k-1}} \hspace{1cm} \mbox{(since $i \leq \frac{1}{p_i}$)}\\
            &=H_k({\p})-p_1\log_k \frac{1}{p_1} +(1-p_1) \log_k(k-1)-0.5 \sum_{j\geq 2: \frac{k^j-1}{k-1}\leq n} p_{\frac{k^j-1}{k-1}}.
        \end{aligned}
    \end{equation*}
    Therefore, (\ref{eq:first}) holds.

    To conclude the proof, it remains  to prove that 
    \begin{equation}
        \label{eq:second}
        L_{\epsilon} \leq H_k({\p}) - \mathcal{H}_k(p_1) + (1-p_1) (\log_k2 (k-1))-0.5 \sum_{j\geq 2: \frac{k^j-1}{k-1}\leq n} p_{\frac{k^j-1}{k-1}}.
    \end{equation}
    By observing that for any $i\geq2$, it holds that
    \begin{equation}\label{eq2}
        p_i \leq \frac{2 (1-p_1)}{i},
    \end{equation}
we obtain:
        \begin{align*}
            L_{\epsilon} &\leq \sum_{i=2}^n p_i\log_k(i (k-1)) - 0.5\!\!\!\!\!\! \sum_{j\geq 2: \frac{k^j-1}{k-1}\leq n}\!\!\! p_{\frac{k^j-1}{k-1}}\\
            &\leq \sum_{i=2}^n p_i\log_k\left(\frac{2(1-p_1)}{p_i} (k-1)\right) - 0.5\!\!\!\!\!\! \sum_{j\geq 2: \frac{k^j-1}{k-1}\leq n}\!\!\!\!\! p_{\frac{k^j-1}{k-1}} \hspace{0.2cm} \mbox{(since from (\ref{eq2}), we have $i\leq\frac{2(1-p_1)}{p_i}$)}\\
            &= H_k({\p}) -p_1\log_k \frac{1}{p_1} +(\log_k 2 + \log_k(1-p_1) + \log_k(k-1))(1-p_1) -0.5\!\!\!\!\!\! \sum_{j\geq 2: \frac{k^j-1}{k-1}\leq n}
            \!\!\!p_{\frac{k^j-1}{k-1}}\\
            &= H_k({\p}) - \mathcal{H}_k(p_1) + (1-p_1) (\log_k2 (k-1))-0.5\!\!\!\!\!\! \!\!\sum_{j\geq 2: \frac{k^j-1}{k-1}\leq n} p_{\frac{k^j-1}{k-1}}.
        \end{align*}
    Therefore, (\ref{eq:second}) holds as well.
    
    \smallskip
    
    From (\ref{eq:first}) and (\ref{eq:second}), since 
    $$\sum_{j\geq 2: \frac{k^j-1}{k-1}\leq n} p_{\frac{k^j-1}{k-1}}\geq0,$$
we obtain 
    \begin{equation*}
        L_{\epsilon} \leq H_k({\p})-p_1\log_k\frac{1}{p_1} +(1-p_1) \log_k(k-1),
    \end{equation*}
    and
    \begin{equation*}
        L_{\epsilon} \leq H_k({\p}) - \mathcal{H}_k(p_1) + (1-p_1) (\log_k 2 (k-1)).
    \end{equation*}
    Now, it is easy to verify that $p_1\log_k\frac{1}{p_1}\geq\mathcal{H}_k(p_1) +(1-p_1) \log_k 2$ for $0<p_1\leq0.5$, proving the Lemma.
\end{proof}

Thanks to the result of Lemma {\ref{lemma:ub_code_wt_space}} and to the formula (\ref{eq:relation_with_eps}), the upper bounds 
obtained above can be used to derive our upper bounds on the average length of optimal prefix-free codes with space, as shown in the following theorems.

\begin{Theorem}
    The average length of an optimal prefix-free code with space $C: \{s_1,\dots,s_n\}\to\{0,\dots,k-1\}^+ \cup \{0,\dots,k-1\}^+\sqcup$ satisfies
    \begin{align}
    \label{eq:th_3}
     L(C) \leq& H_k({\p})+\log_k(k-1) + \sum_{i=1}^{\lfloor \log_k \lceil \frac{n-1}{k}\rceil\rfloor} p_{\frac{k^i-1}{k-1}}\nonumber\\
     &+{\sum_{i=1}^{\frac{k^{{h}-1}-1}{k-1}-1} \!\!\!\!p_i  + \sum_{i=\frac{k^{{h}}+k^{{h}-1}-2}{k-1} -\lceil n/k\rceil}^{\frac{k^{{h}}-1}{k-1}-1}\!\!\!\!\!\!\!\!\!\!\!\!\!\! p_i }\\
     \leq& H_k({\p})+\log_k(k-1) + \sum_{i=1}^{\lfloor \log_k \lceil \frac{n-1}{k}\rceil\rfloor} p_{\frac{k^i-1}{k-1}} + \sum_{i=1}^{\lceil \frac{n}{k}\rceil-1}
     p_i,
\end{align}
where {$h=\lceil\log_k (n-\lceil n/k\rceil)\rceil$}.
\end{Theorem}
\begin{proof}
  From Lemma {\ref{lemma:ub_code_wt_space}} and the formula (\ref{eq:relation_with_eps}), we have
    \begin{align}
        L(C) \leq& L_{\epsilon} + \sum_{i=1}^{\lfloor \log_k \lceil \frac{n-1}{k}\rceil\rfloor} p_{\frac{k^i-1}{k-1}}\nonumber\\
        &+{\sum_{i=1}^{\frac{k^{{h}-1}-1}{k-1}-1} \!\!\!\!p_i  + \sum_{i=\frac{k^{{h}}+k^{{h}-1}-2}{k-1} -\lceil n/k\rceil}^{\frac{k^{{h}}-1}{k-1}-1}\!\!\!\!\!\!\!\!\!\!\!\!\!\! p_i 
        }.\label{eq:length_of_optimal_prefix_free_3}
    \end{align}
    By applying the upper bound (\ref{eq:lemma4}) on $L_{\epsilon}$ of Lemma \ref{Wyner_gen} to (\ref{eq:length_of_optimal_prefix_free_3}), we obtain (\ref{eq:th_3}).  
\end{proof}

\begin{Theorem}
    The average length of an optimal prefix-free code with space $C: \{s_1,\dots,s_n\}\to\{0,\dots,k-1\}^+ \cup \{0,\dots,k-1\}^+\sqcup$ satisfies
    \begin{align}
    \label{eq:th_4}
    L(C) \leq 
        \begin{cases}
        H_k({\p}) -p_1\log_k \frac{1}{p_1} + (1-p_1) \log_k(k-1)\\
        +\sum_{i=1}^{\lceil \frac{n}{k}\rceil-1} p_i + \sum_{i=1}^{\lfloor \log_k \lceil \frac{n-1}{k}\rceil\rfloor} p_{\frac{k^i-1}{k-1}} \hspace{2cm}\text{ if } 0<p_1\leq 0.5,
        \vspace{0.5cm}\\
        H_k({\p}) -\mathcal{H}_k(p_1) + (1-p_1) \log_k 2 (k-1)\\
        +\sum_{i=1}^{\lceil \frac{n}{k}\rceil-1} p_i + \sum_{i=1}^{\lfloor \log_k \lceil \frac{n-1}{k}\rceil\rfloor} p_{\frac{k^i-1}{k-1}}\hspace{2cm}\text{ if } 0.5<p_1\leq 1.
    \end{cases}
\end{align}
\end{Theorem}
\begin{proof}
     From Lemma {\ref{lemma:ub_code_wt_space}} and the formula (\ref{eq:relation_with_eps}) and by recalling that  {$h=\lceil\log_k (n-\lceil n/k\rceil)\rceil$}, we have
    \begin{align}
        L(C) \leq& L_{\epsilon} + \sum_{i=1}^{\lfloor \log_k \lceil \frac{n-1}{k}\rceil\rfloor} p_{\frac{k^i-1}{k-1}}\nonumber\\
        &+{\sum_{i=1}^{\frac{k^{{h}-1}-1}{k-1}-1} \!\!\!\!p_i  + \sum_{i=\frac{k^{{h}}+k^{{h}-1}-2}{k-1} -\lceil n/k\rceil}^{\frac{k^{{h}}-1}{k-1}-1}\!\!\!\!\!\!\!\!\!\!\!\!\!\! p_i }\nonumber\\
        \leq& L_{\epsilon} + \sum_{i=1}^{\lceil\frac{n}{k}\rceil-1} p_i+ \sum_{i=1}^{\lfloor \log_k \lceil \frac{n-1}{k}\rceil\rfloor} p_{\frac{k^i-1}{k-1}}.\label{eq:length_of_optimal_prefix_free_4}
    \end{align}
    We apply the upper bound 
    (\ref{eq:Blundo_gen}) on $L_{\epsilon}$ of Lemma \ref{Blundo_gen} to (\ref{eq:length_of_optimal_prefix_free_4}). That gives us  (\ref{eq:th_4}).  
\end{proof}

{
\begin{Remark}
    One can  estimate how much the average length 
    of optimal prefix-free codes ending with a space
     differs from the minimum average length of
     \emph{unrestricted} optimal prefix-free codes on the alphabet
     $\{0, 1, \ldots, k-1, \sqcup\}$, that is optimal 
     prefix-free codes in which the special symbol $\sqcup$ is not
     constrained to appear at the end of the codewords, only.
     
     Let $S=\{s_1,\ldots,s_n\}$ be the set of source symbols and 
    $\p=(p_1,\ldots , p_n)$ be a probability distribution on $S$. Let us denote by $C_{\sqcup}: \{s_1,\dots,s_n\}\to\{0,\dots,k-1\}^+ \cup \{0,\dots,k-1\}^+\sqcup$ an optimal prefix-free code ending with a space for $S$ and by $C:\{s_1,\dots,s_n\}\to\{0,\dots,k-1,\sqcup \}^+$  an optimal prefix-free code without the restriction of 
    where the space can occur. 
    Clearly, $L(C_{\sqcup})<H_k(\p)+1$, since the more constrained optimal 
    code $C': \{s_1,\dots,s_n\}\to\{0,\dots,k-1\}^+$ has an average length
    less than $H_k(\p)+1$. Therefore,
    \begin{align*}
        L(C_{\sqcup}) -L(C) <& H_k(\p)+1 - L(C) \\
        \leq & H_k(\p)+1 - H_{k+1}(\p)
       \qquad \mbox{(since $L(C)\geq H_{k+1}(\p)$)}\\
        =& H_{k}(\p) \left(1-\frac{1}{\log_k (k+1)}\right)+ 1.
    \end{align*}
    Since $\lim_{k \to \infty}\log_k (k+1) = 1$, we have that, as the cardinality of the code alphabet increases, the constraint that the space can appear only at the end
    of codewords 
    becomes less and less influential. 
\end{Remark}
}

\section{Concluding Remarks}
{
In this paper, 
we have introduced  
the class of prefix-free codes where a specific symbol
(referred to as a ``space") can only appear at the end of codewords. 
We have proposed 
 a linear-time algorithm to construct 
``almost"-optimal codes with this characteristic, 
and we have shown that their  average length  is 
at most one unit longer than the \textit{minimum} 
average length of any prefix-free code in which the space can appear 
only at the end of codewords. We
have proven this result by highlighting
a connection between our type of codes and the 
well-known class of one-to-one codes. We have also provided upper and lower limits of the average length of optimal prefix-free codes ending with a space, expressed in 
terms of the source entropy and
the cardinality of the code alphabet.

We leave open the problem of providing an efficient algorithm
to construct \textit{optimal} prefix-free codes ending with a space. 
It seems that there is no easy way to modify the classical 
Huffman greedy algorithm to solve our problem. 
It is possible that the more
powerful dynamic programming approach could 
be useful to provide an optimal solution
to the problem, {as done in \cite{CG} 
for optimal binary codes ending with ones}.  This will be the subject of future investigations.
}

\vspace{6pt} 





\reftitle{References}

\end{document}